\documentclass[12pt]{article}
\usepackage{amssymb}
\usepackage{amsmath}
\usepackage{bbm}
\usepackage{graphicx,psfrag,epsf}
\usepackage{enumerate}
\usepackage[titletoc, title]{appendix}
\usepackage{url} 

\usepackage[colorinlistoftodos, textwidth=31mm, textsize=12pt]{todonotes}
\usepackage{hyperref}

\DeclareMathOperator*{\argmin}{arg\,min}

\usepackage{caption}
\usepackage{geometry}
\usepackage{float}
\usepackage{placeins}
\usepackage{url}
\usepackage{listings}
\usepackage{enumitem}
\usepackage{booktabs}
\usepackage[english]{babel}
\newcommand{\blind}{0}

\begin{document}

\def\spacingset#1{\renewcommand{\baselinestretch}%
{#1}\small\normalsize} \spacingset{1}


\if0\blind
{
  \title{\bf Subgroup identification in dose-finding trials via model-based recursive partitioning}
  \author{Marius Thomas\\
    Novartis Pharma AG, Basel, Switzerland\\
    \vspace{0.05cm}\\
    Bj\"orn Bornkamp\\ 
    Novartis Pharma AG, Basel, Switzerland\\
   \vspace{0.05cm}\\
    Heidi Seibold\\
    University of Zurich, Zurich, Switzerland}
  \date{}
  \maketitle
} \fi

\if1\blind
{
  \bigskip
  \bigskip
  \bigskip
  \begin{center}
    {\LARGE\bf Subgroup identification in dose-finding trials via model-based recursive partitioning}
\end{center}
  \medskip
} \fi

\bigskip
\begin{abstract}
An important task in early phase drug development is to identify patients, which respond
better or worse to an experimental treatment. While a variety of different subgroup
identification methods have been developed for the situation of trials that study
an experimental treatment and control, much less work has 
been done in the situation when patients are randomized to different dose groups.
In this article we propose new strategies to perform subgroup analyses in 
dose-finding trials and discuss the challenges, which arise in this new setting.
We consider model-based recursive partitioning, which has recently been applied to subgroup
identification in two arm trials, as a promising method to tackle
these challenges and assess its viability using a real trial example and simulations.
Our results show that model-based recursive partitioning can be used 
to identify subgroups of patients with different dose-response curves
and improves estimation of treatment effects and minimum effective doses, when
heterogeneity among patients is present.
\end{abstract}

\noindent%
\vfill

\noindent
\small{This is the pre-peer reviewed version of the following article:\\
Thomas, M., Bornkamp, B., \& Seibold, H. (2018). Subgroup identification in dose-finding trials via model-based recursive partitioning. \textit{Statistics in medicine}, 37(10), 1608-1624, \\
which has been published in final form at \url{https://doi.org/10.1002/sim.7594}. This article may be used for non-commercial
purposes in accordance with Wiley Terms and Conditions for Self-Archiving.}
 
\newpage
\spacingset{1.45} 

\section{Introduction}
The identification of subgroups (defined in terms of baseline
covariates) with a modified response to a treatment is an important,
but also challenging task in drug development. Firstly the
identification task itself is not trivial: Covariates may act on the
response independent of any treatment (prognostic covariates), but
usually one is interested in covariates modifying the response to the
specific treatment administered (predictive covariates). In addition,
the treatment effect may be defined in terms of a non-trivial function
of the covariates. Further statistical issues include multiplicity,
bias in treatment effect estimates in selected subgroups (due to the
selection) and sample sizes that are typically too low to detect
relevant differences between subgroups. A high-level overview of the
involved statistical challenges but also opportunities is given in
\cite{rube:shen:2015}.

The development of computational and statistical tools to identify
subgroups/covariates leading to a differential response has been a
major focus of statistical research in recent years. Due to their
ability to handle high-order interactions and their good
interpretability, many of the proposed approaches employ tree-based
partitionings of the overall trial data \cite{su:tsai:wang:2009, fost:2011, 
lipk:dmit:denn:2011, duss:vanm:2014, loh:he:man:2015}. 
Other statistical approaches to the problem include
Bayesian models \cite{berg:wang:shen:2014} and penalized regression
coupled with a transformation of the covariates \cite{tian:ash:gent:2014}.  
A recent overview paper on the topic of
subgroup identification is \cite{lipk:dmit:agos:2017}.

Most of these methods have been developed in the context of clinical
trials that compare a new treatment against a control. However studies
with more than one dose of the active treatment are also common in
clinical trials conducted in late-stage development. This is obviously
true for Phase II dose-finding trials (see \cite{born:2017, ting:2006}
for an overview), but also in Phase III trials sometimes two or three
doses are studied. When patients are administered different doses of
the new treatment and a subgroup search is performed, this search has
to be adjusted for the fact that patients received different doses:
One patient might respond better compared to another patient because
they differ in the dose received but not because of their differing
baseline characteristics.

One way to approach this problem is to assume a functional relationship
between dose and response to account for dose, but allow that this
relationship can differ across subgroups. 
One can for example assume a the $E_{max}$
function, a standard parametric model in dose-response analyses 
\cite{born:2017, thom:swee:soma:2014}, which leads to a function
non-linear in its parameters. Alternatively a spline function might be
used. Another alternative would be to use a model that
just describes the treatment effect at the observed doses and makes
no assumptions about the functional replationship.  The idea
in all three cases would then be that within each subgroup the same
type of model is fitted, but the model parameters would vary between
subgroups.

\begin{figure}[h!]
\centering
\includegraphics[width=0.85\textwidth]{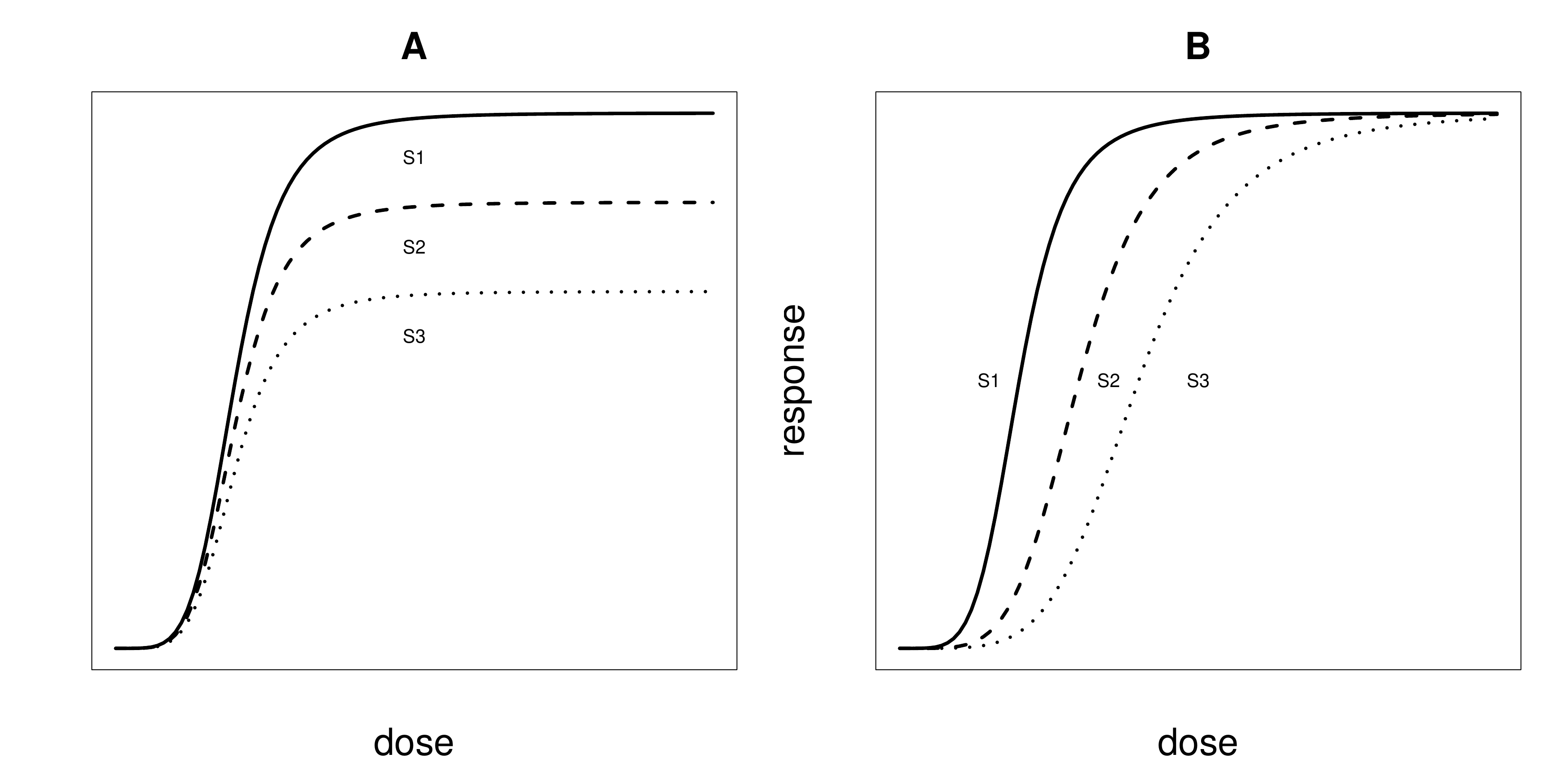}
\caption{Example of different dose-response shapes in three exemplary
  subgroups S1-S3. A shows subgroups with
  different maximum treatment effects, 
B shows an example of subgroups of patients
  potentially requiring different doses, }
\label{fig:ex_subs}
\end{figure}

In the setting of dose-response analyses the consequences of finding
subgroups is different than in standard subgroup analyses, because
subgroups might differ not only in their treatment effect but more
generally in terms of the shape of the dose-response curve. One might,
for example, identify subgroups that have an increased or decreased
treatment effect (such as in Figure \ref{fig:ex_subs}A). But one might
also identify subgroups requiring a different dose to achieve a
desired treatment effect (such as in Figure \ref{fig:ex_subs}B).

In this paper we want to propose the use of
model-based recursive partitioning in this setting. 
This approach has originally been proposed in \cite{zeil:hoth:horn:2008} 
and applied to subgroup analyses for two-armed trials 
in \cite{seib:zeil:hoth:2016}. This algorithm
splits the overall trial population recursively into subgroups
(defined by baseline covariates), so that each group has a homogeneous
dose-response relationship. The advantages of this approach compared
to other approaches for recursive partitioning are that (i) it is
easily adaptable to different statistical models (such as different
endpoint distributions and also nonlinear models), (ii) it contains an
easily interpretable stopping criterion to control the complexity of
the tree and (iii) allows to restrict the splitting to specific model
parameters, which can be helpful to distinguish between prognostic and
predictive effects.  In addition the methodology is implemented in the
publicly available R package \textit{partykit} \cite{hoth:zeil:2015}.
The purpose of this paper is to investigate the suitability of this 
method in the context of dose-finding trials.

The paper is structured as follows:
In the next section we will introduce a motivating example dose-finding
study. In Section \ref{sec:stat-meth} we will first discuss the considered
dose-response models and then introduce model based
recursive partitioning in this context. Section
\ref{sec:simulation-study} contains a simulation study evaluating the
properties of the proposed method and comparing different models. In
Section \ref{sec:ex-analysis} the trial example will be revisited and
analysed to illustrate the methodology on a concrete data example.
Conclusions and some discussion are presented in Section
\ref{sec:discussionconclusion}.

\section{A motivating example}
\label{sec:motivating-example}

Exploratory subgroup analyses are performed in many stages of clinical
drug development. In Phase II these analyses are performed to identify
whether any and/or which baseline covariates modify the treatment
effect. The result of these analyses will inform decisions regarding
the further development of the drug, for example how subsequent
clinical trials are designed.

As an example we consider data from a dose-finding trial conducted to
assess the efficacy of a new treatment for an inflammatory disease.
For reasons of confidentiality all variable names are non-descriptive
and all continuous variables have been rescaled to have mean 0 and standard
deviation 1.  We have complete
data from 270 patients, which were distributed across 4 arms,
receiving either a placebo ($n = 75$) or the new drug at dose levels
25 ($n= 54$), 50 ($n = 62$) and 100 ($n= 79$). The primary endpoint is
the change from baseline in a continuous variable. Additionally
baseline measurements of 10 covariates -- 6 of which are categorical
and 4 of which are continuous -- are available for each patient.

\begin{figure}[h!]
\centering
\includegraphics[width=0.7\textwidth]{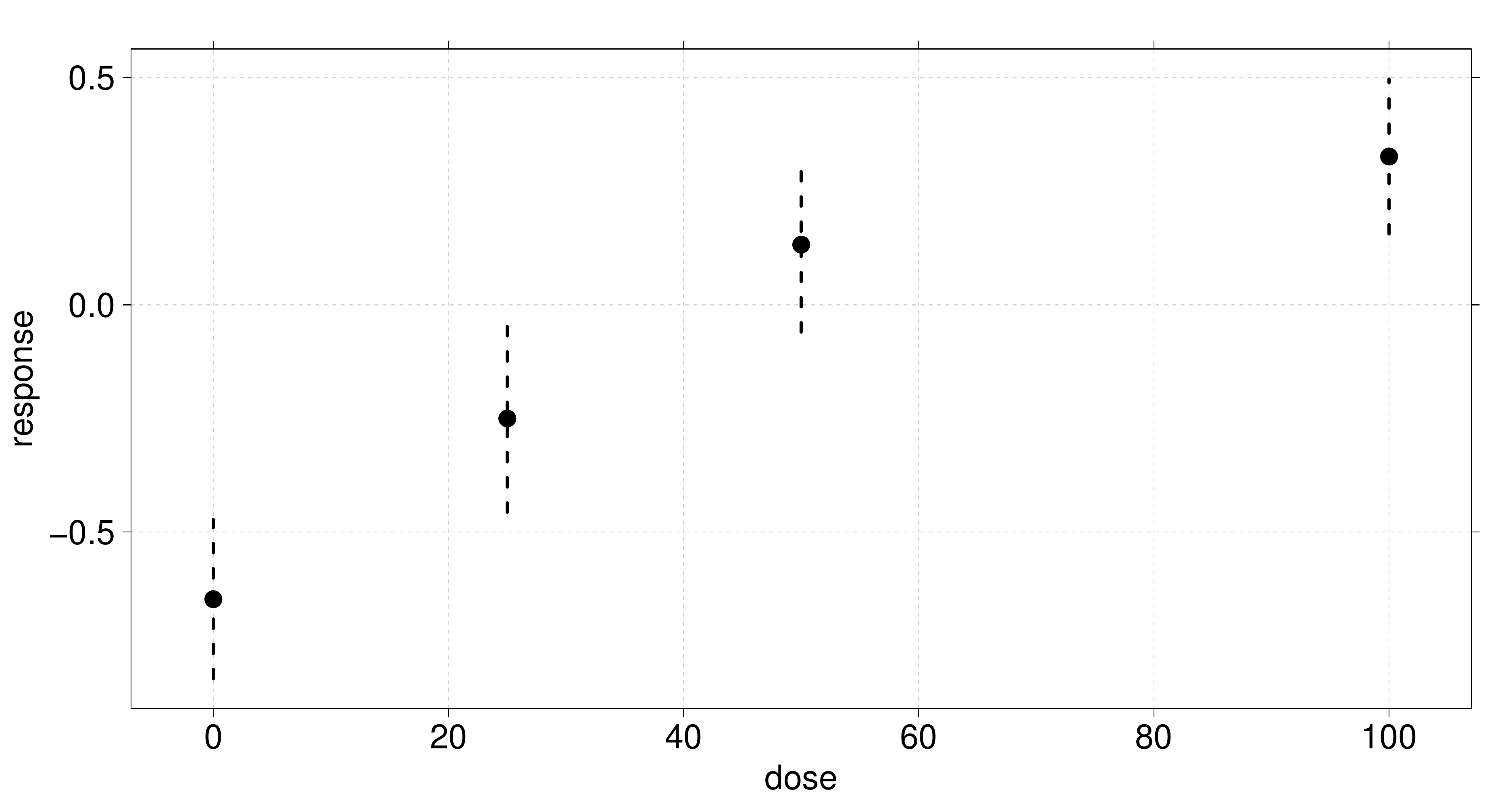}
\caption{Mean responses and $90\%$-confidence intervals for the dose-finding trial example.}
\label{fig:ex_emax}
\end{figure}

The mean responses at the dose levels in the trials along with the
confidence intervals are shown in Figure \ref{fig:ex_emax},
which suggest a clear dose-response effect. Still there is interest in
further investigation of whether there is a subgroup with differential
treatment effect. This analysis and its results are discussed in
Section \ref{sec:ex-analysis}, using the methodology developed in Section \ref{sec:stat-meth}.

\section{Statistical methodology}
\label{sec:stat-meth}

In this section we will first introduce the dose-response models we
consider. Even though we focus on normally distributed endpoints in the
rest of the article, we will introduce the models in a generalized form,
which also allows for other commonly encountered outcome types in
clinical trials, such as binary, count or time-to event data.

In the second part of this section we will then introduce the
model-based recursive partitioning method in the context of
dose-response modeling. A more detailed discussion of the algorithm
can be found in \cite{zeil:hoth:horn:2008}.

\subsection{Model specification} 
\label{sec:model-spec}

We consider the situation of a clinical trial with $n$ patients, that
receive doses $d_1,...,d_n$ of a new treatment at $l$ dose levels, so
that $d_i \in \{\tilde{d_1},...,\tilde{d_l}\}$, where the lowest level
is a placebo $\tilde{d_1} = 0$. We observe responses $y_1,...,y_n$,
which can be related either to efficacy or safety of the drug in
question. Typically a small set of additional baseline covariates
$\boldsymbol{x_i}$ of dimension $K$ for each patient $i$, are measured
and adjusted for in the analysis, examples are the baseline value of
the outcome (if change from baseline is used) or covariates like
region, center or other stratification variables.

Additionally we assume, that $E(y_i) =
\mu_i = g^{-1}(\eta_i)$. Here $g$ denotes a link function
that maps from the space of the response variable to $\mathbb{R}$ and
$\eta_i$ is a (possibly non-linear) predictor such that 
\begin{equation}
\label{eq:model-specification}
\eta_i = \beta_0 + \Delta(d_i, \boldsymbol{\theta}) + \boldsymbol{\gamma}^{'}\boldsymbol{x_i}. 
\end{equation}
$\beta_0$ is an intercept describing the response under placebo and
$\Delta(d_i, \boldsymbol{\theta})$ is a dose-response function with
parameter vector $\boldsymbol{\theta}$ (with
$\Delta(0, \boldsymbol{\theta})=0$), which describes the treatment
effect in dependence of the dose level. Additional covariate main
effects are modeled in $\boldsymbol{\gamma}$.

In total we obtain a model
$m((\boldsymbol{y}, \boldsymbol{d}, \boldsymbol{X}),
\boldsymbol{\varphi})$
with $\boldsymbol{y}=(y_1,...,y_n)$, $\boldsymbol{d}=(d_1,...,d_n)$,
$\boldsymbol{X}=(\boldsymbol{x}_1,...,\boldsymbol{x}_n)$ and
$\boldsymbol{\varphi} = (\beta_0, \boldsymbol{\theta},
\boldsymbol{\gamma}, \sigma)$,
where $\sigma$ is a nuisance parameter, as for example the standard
deviation of the response in a Gaussian GLM.  An estimate for
$\boldsymbol{\varphi}$ can be derived as
\begin{equation*}
\hat{\boldsymbol{\varphi}} = \argmin\limits_{\boldsymbol{\varphi}}\sum\limits_{i = 1}^{n}\Psi((y_i, d_i, \boldsymbol{x_i}), 
\boldsymbol{\varphi}),
\end{equation*}
by minimizing the objective function $\Psi$ corresponding to model
$m$, for example the log-likelihood. In what follows we will only
consider normally distributed outcomes, so that this reduces to the
residual sum of squares in our setting. Finding
the minimum above is equivalent to solving
\begin{equation*}
\sum\limits_{i = 1}^n \frac{\partial\Psi((y_i, d_i, \boldsymbol{x_i}), 
\boldsymbol{\varphi})}{\partial\boldsymbol{\varphi}} = \sum\limits_{i = 1}^{n}\psi((y_i, d_i, \boldsymbol{x_i}), 
\boldsymbol{\varphi}) = 0,
\end{equation*}
where $\psi$ is the score function.

We will now give an overview over the different functional forms of
$\Delta$, which will be used to model dose-response relationships in the
remainder of this article. These forms of $\Delta$ contain non-linear
and linear models and show varying degrees of flexibility in regards to
the dose-response shapes, which they can fit. Table \ref{tab:mod} shows
a summary.

\begin{table}[h]

  \caption{Functional forms of $\Delta$ used to model dose-response. $b_1, b_2, b_3$
  denote basis functions of a B-spline with degree 2 and one inner knot at the median of the dose levels.}
  \centering
   \begin{tabular}{ccc}
    \toprule
    Model & $\Delta(., \boldsymbol{\theta})$ & Number of parameters\\
    \midrule
    $E_{max}$ & $\theta_1 \frac{d}{\theta_2 + d}$ & 2\\
    B-Spline &  $\theta_1b_1(d) + \theta_2b_2(d) + \theta_3b_3(d)$  & 3\\
    Means &  $\Delta(\tilde{d_j}, \boldsymbol{\theta})=\theta_j$, $j=1,...,l$ & $l - 1$\\
    \bottomrule
 \end{tabular}

 \label{tab:mod}
\end{table}

The $E_{max}$ model is commonly used to model plateauing, monotonic
dose-response functions \cite{born:2017}. It has been shown to be
adequate in a wide variety of real dose-response situations, see
\cite{thom:swee:soma:2014}, and is, as many other dose-response
models, non-linear in its parameters. It can be derived from
pharmacological principles \cite{kael:2007} and its parameters have a
direct interpretation, as $\theta_1$ represents the maximum treatment
effect, which is reached as the dose goes to infinity, and $\theta_2$
represents the dose required to receive $50\%$ of this maximum effect.

To relax the parametric assumptions, we also consider a spline model,
which is linear in the parameters and uses transformations of the dose
variable as the independent variable in the model. As only a limited
number of dose-levels is typically used in dose-finding trials, we use
quadratic B-splines with only one knot at the median of the active
dose levels. This model can fit a large number of possible
dose-response shapes well, but requires the estimation of one more
parameter than the $E_{max}$ model.

The last ``Means'' model estimates the treatment effects
independently, without assuming a relationship among the doses. It
does not make any assumptions about the underlying dose-response
relationship and treats the different dose levels as independent
treatments. This allows for additional flexibility in the possible
dose-response relationship, but can possibly overfit the data and lead
to biologically implausible estimated dose-response relationships. In
addition this model does not allow to predict the dose-response effect
beyond the actually observed doses.  
An interpolation method, for example using splines, 
is required to obtain predicted responses for dose levels 
which lie between the observed dose levels.


\subsection{Model-based recursive partitioning}
\label{sec:meth-mob}

Primary, pre-specified analyses of clinical trials typically assume
that the treatment effect is homogeneous across the population
studied. Baseline covariates (as in Equation
\ref{eq:model-specification}) are typically included only if they are
considered important already at the design stage of the
trial. Nevertheless additional baseline covariates $\boldsymbol{Z}$
might have been measured and it is of interest to evaluate the effect
that these might have on the dose-response model.


This can be achieved with model-based recursive partitioning (or
\textit{mob} for short). This approach applies a parametric model
and allows for the parameters to depend on certain baseline covariates
(e.g. biomarkers) $\boldsymbol{Z}$. We can thus rewrite the model as
$m\left((\boldsymbol{y}, \boldsymbol{d}, \boldsymbol{X}),
  \boldsymbol{\varphi}(\boldsymbol{Z})\right)$.
This is achieved by estimating separate models for different
subgroups. These subgroups are found through an algorithm, which
recursively tries to detect, if there are covariate effects on the
parameters of model $m$. If covariate effects are detected, the
algorithm will split the patients into subgroups using these
covariates. Then in each of these subgroups models are estimated
separately.

To detect covariate effects the model-based recursive partitioning algorithm
makes use of tests for parameter instability in model $m$. Parameter
instability can be discovered by testing for independence between the
partial scores and covariates $\boldsymbol{z^{(1)}}, \dots, \boldsymbol{z^{(J)}}$, 
i.e.
\begin{align}\label{eq:null}
\psi_{\varphi_p}((\boldsymbol{y}, \boldsymbol{d}, \boldsymbol{X}),
\hat{\boldsymbol{\varphi}}) \quad\bot\quad \boldsymbol{z^{(j)}}, \quad j = 1, \dots, J,
\quad p = 1, \dots, P
\end{align}
with $J$ being the number of partitioning covariates and $P$ the
number of model parameters. The partial score $\psi_{\varphi_p}$ is
the partial derivative of the objective function with respect to the
model parameter $\varphi_p$ respectively. For a detailed
discussion of the algorithm and the instability tests used, see
\cite{zeil:hoth:horn:2008} and \cite{zeil:horn:2007}.

If the overall null
hypothesis of no instability (for any of the parameters) is not rejected, 
we assume no (further) subgroups.
If it is rejected, the variable corresponding to the smallest $p$-value is
chosen as split variable. The subgroups are formed based on this split variable
using a binary split. 
If there are multiple possible splits over the variable, the split, which minimizes the
objective function of the model in the two resulting subgroups is chosen. 
Then new models are estimated in each subgroup and the parameters of
each model are again tested for instability. New subgroups are formed until
the overall null hypothesis can no longer be rejected or other stop criteria
come into effect (for example the minimum subgroup size is reached).

In each step of the algorithm (for each split) $J \times P$ null
hypotheses (see Equation \ref{eq:null}) are tested. To adjust
for the fact that multiple tests are performed 
a Bonferroni correction is used.

In addition one might be especially interested in specific
parameters. For example in the dose-response models the effects
$\boldsymbol{\gamma}$ of baseline covariates and nuisance parameters
${\sigma}$ may only be of secondary interest. Then it is possible to
restrict the instability tests on partial scores with respect to
parameters $\beta_0$ and $\boldsymbol{\theta}$.  For the subgroup
analyses we consider here, one might even go further and restrict the
splitting only to $\boldsymbol{\theta}$, since only these
parameters impact the treatment effect.

\section{Simulation study}
\label{sec:simulation-study}

This section will show the results of simulations to evaluate the
performance of the previously described methods in simulation
scenarios.  Our simulation scenarios aim to represent the situation of
a Phase II dose-finding trial, for which exploratory subgroup analyses
are regularly conducted. The main characteristics of the study we
simulate, e.g. the general form of the dose-response curve, the dose
levels used and the standard error are based on a study to investigate
the efficacy and safety of glycopyrronium bromide in COPD
patients. The summary statistics from this study can be found on
\texttt{clinicaltrials.gov} under NCT00501852.

We simulate clinical trials with $n=250$ patients,
which are equally distributed across 4 active doses (12.5, 25, 50, 100)
and a placebo, resulting in 5 dose levels in total with 50 patients
each. We generate a vector of baseline covariates for each patient $i$
as $\boldsymbol{z_i} \sim N(\boldsymbol{0}, \boldsymbol{I_{10}})$. We generate data
from an $E_{max}$ model of the form
\begin{equation}
\begin{split}
y_i\sim N(\mu_i, \sigma^2) \text{ i.i.d},\\
\mu_i = \beta_0(\boldsymbol{z_i}) + \theta_1(\boldsymbol{z_i})\frac{d_i}{\theta_2(\boldsymbol{z_i}) + d} \text{ for i=1,...n,}
\end{split}
\label{eq:datmodel}
\end{equation}
For this model the parameters describing placebo response ($\beta_0$), 
maximum effect ($\theta_1$)
and the dose giving 50\% of the response ($\theta_2$) depend on a
patient's baseline covariates.  The ratio of effect size to noise can
be controlled through $\sigma$.  Unless explicitly stated otherwise
$\sigma$ is always set to 0.12 in the following simulations.

\begin{table}[h]
\caption{Base cases for simulation study. Here $I_j = I(\boldsymbol{z^{(j)}} > 0)$. The
   rightmost column gives the number of groups, e.g. the size of
   a partition needed to achieve completely homogeneous groups of patients
   with regards to the parameters of the model.}
\centering
\footnotesize
  \[
   \begin{array}{lllll}\toprule
   \text{Case} & \beta_0(\boldsymbol{z}) & \theta_1(\boldsymbol{z}) & \theta_{2}(\boldsymbol{z}) & \text{number of groups} \\\midrule
    \text{1 (Null)} & 1.2 & 0.17 & 18 & 1\\
    \text{2 ($\beta_0$)} & 1.2 + 0.1I_1 + 0.1I_3 & 0.17 & 18 & 4\\
    \text{3 ($\theta_{1}$)} & 1.2 & 0.17 - 0.17I_1 + 0.17I_2 & 18 & 4\\
    \text{4 ($\theta_{2}$)} & 1.2 & 0.17 & 18\cdot 0.2^{I_1}\cdot 5^{I_2} & 4\\
    \text{5 (all)} & 1.2 + 0.1I_1 + 0.1I_3  & 0.17 + 0.17I_1I_2 - 0.17(1 - I_1)(1 - I_2) & 18 \cdot 0.2^{I_1} & 8\\\bottomrule
  \end{array}
  \]
  
 \label{tab:cases}
\end{table}

We consider five cases for our simulation study, which include a null
case with no covariate effects, three cases with covariate effects on
only one of the parameters of the model and one case with
covariate effects on all parameters. For details see Table \ref{tab:cases}.
With the 5 considered cases we simulate scenarios, where there are:
\begin{enumerate}[label=\textbf{\arabic*}]
\item no subgroups;
\item subgroups with differences in placebo response (change in $\beta_0$); 
\item subgroups with a doubled treatment effect, while others have none (change in $\theta_1$);
\item subgroups, for which the dose-response curve is very steep,
and plateaus near the lower end of the dose range,
while for others the curve is very shallow and the plateau 
is not reached at the maximum dose (change in $\theta_2$);
\item a combination of 2-4.
\end{enumerate}

The following simulations can be divided into two parts with different
objectives. 
The first part tries to evaluate how well the \textit{mob} procedure
resolves the bias-variance trade-off in data partitioning: A global
model fitted to all data might be biased, but will have smaller
variability in parameter estimates, compared to a model that uses the
true partitions. Using the true partitions will lead to unbiased
parameter estimates but larger variability in parameter estimates (due
to smaller sample size in each subgroup). The \textit{mob} procedure
can be considered as an adaptive procedure, as it decides on the model
complexity adaptively (\textit{i.e.}  it might also fit a global model
if no split is selected in the first step). These first simulations
therefore compare \textit{mob} to these two extreme cases (global
model and the model using the true partitions). Their results are
discussed in Section \ref{sec:results-fit}.

The second part of the simulations is concerned with evaluating the
performance of the algorithm in regards to subgroup analyses.  In
these simulations, which are discussed in Section
\ref{sec:results-id}, the subgroup identification and estimation
performance is evaluated.

We use the function \textit{mob} in the R package \textit{partykit}
together with our own user-defined fitting function, which can be found
in the appendix. We focus on the
case of a normally distributed response variable and include no
additional baseline covariates in the models.  The trees use models of
the form
\begin{gather*}
y_i\sim N(\mu_i, \sigma^2) \text{ i.i.d},\\
\mu_i = \beta_0 + \Delta(d_i, \boldsymbol{\theta}) \text{ for i=1,...,n,}
\end{gather*}
where $\Delta$ has one of the functional forms shown in Table
\ref{tab:mod}, the corresponding methods will be called mobEmax, mobSpline and
mobMeans in what follows. We will use the algorithm described in
\ref{sec:meth-mob} to detect instabilities over the partitioning
variables $\boldsymbol{z}$.  Two different approaches for selection of
variables for the splits are used. In one approach instabilities in
both $\beta_0$ and $\boldsymbol{\theta}$ are tested (denoted as
``unrestricted" in what follows), in the other testing is restricted
to the parameters in $\boldsymbol{\theta}$, which affect the
dose-response function (denoted as ``restricted'' in what
follows). This is implemented via the \textit{parm} argument in
\textit{mob}.

As additional arguments for the \textit{mob} function we use control
parameters of \textit{alpha = 0.1, minsize = 20} and \textit{maxdepth
  = 4}, which respectively control the significance level used for the
instability tests at each split, the minimum size of the subgroups and
the maximum depth of the tree. 

\subsection{Improvement in model fit}
\label{sec:results-fit}
Partitioning the data when covariate effects are present should
improve the model fit on independent test data, compared to a global
model. Evaluating if there is a benefit in partitioning the data and
if the partitioning algorithm can reliably find these partitions
should be the first step, when assessing the properties of the
method. For this purpose we generate training sets of size $n=250$ and
test sets of size $n=10000$ from the model in Equation
(\ref{eq:datmodel}). We use the training set to fit the model and
calculate the log-likelihood of the test set data under the fitted
model. We compare three models: the global model, which fits one model
for all patients, the model partitioned by \textit{mob} and a model
fit on the true partition.

\begin{figure}[h!]
\centering
\includegraphics[width=0.9\textwidth]{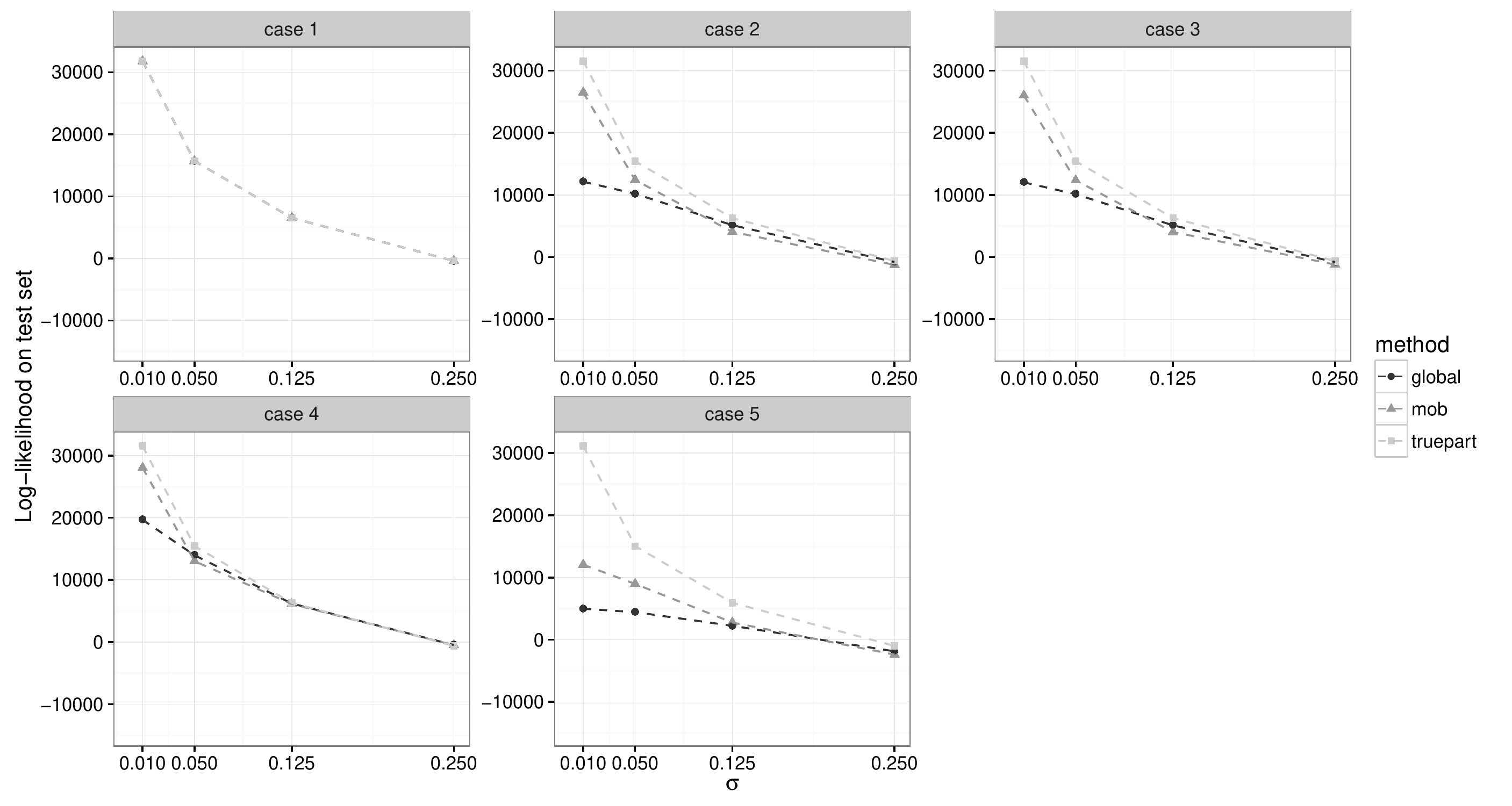}
\caption{Median log-likelihood of test set data of size $n=10000$ under
$E_{max}$ models fit on training data of size $n=250$ for all simulation cases
and varying standard deviations $\sigma$. Global refers to one model 
fit for all patients, \textit{mob} to model-based recursive partitioning and truepart
to models fit on the true partition.}
\label{fig:ll-emax}
\end{figure}

Figure \ref{fig:ll-emax} shows the median of the log-likelihood on the test
sets over 5000 simulations for the cases in Table \ref{tab:cases} depending on
standard deviations $\sigma$, when using $E_{max}$ models.
The plot visualizes the trade-off between bias and variance. For small
$\sigma$ partitioning the data clearly leads to a better model fit,
since a larger fraction of the variance is explained by covariate
effects on the model parameters rather than the error variance and
refraining from partitioning the data leads to biased inference for
some patients.  As $\sigma$ increases, the benefit of partitioning the
data decreases. Partitioning the data and fitting models separately on
a small group of patients leads to parameter estimates that are too
variable, when the data are noisy (even when using the true
partition). 

As \textit{mob} adaptively decides if the data should be partitioned
the model fit on the test set is usually somewhere between the global
model and using the true partition (which is of course unknown in
reality). The only exception for this seem to be the scenarios with
$\sigma = 0.125$ and $\sigma = 0.25$ and under case 2, 3 and 5, where
\textit{mob} sometimes shows a slightly worse fit than both the global
model and the model using the true partition.  In the scenarios with
smaller $\sigma$ \textit{mob} improves the model fit over the global
model. In scenarios with more noise \textit{mob} often refrains from
partitioning the data and therefore shows a similar fit as the global
model.


\subsection{Subgroup identification and estimation of dose-response curve}
\label{sec:results-id}

In this section performance metrics relevant for subgroup analyses
will be discussed in more detail.

\paragraph{Identification of correct covariates}
Identifying the covariates, that interact with the treatment
is a main interest of subgroup analyses. Based on these covariates
subgroups of patients with differential responses to the treatment
can be defined. In this section we will therefore discuss
the results of simulations, which aim to evaluate the capability of
the method to partition over the correct baseline covariates.
Additionally we investigate the rate of false positive discoveries,
when there are no interactions between covariates and treatment.
For these purposes we track the structure of the tree, e.g.
for each trial simulation and each of the 10 covariates, we track if
the covariate is used as a split variable in the corresponding mob tree.
Table \ref{tab:id1} shows the results averaged over 5000 simulations.

\begin{table}[ht]
\caption{Relative frequency of covariates appearing in the trees for different \textit{mob} models with and without restriction to treatment effect parameters over 5000 simulations.
Covariates in bold are interacting with the treatment effect. z1 and z3 are prognostic for
Cases 2 and 5.}
\centering
\tiny
\begin{tabular}{rr|ccccc|ccccc}
 & &&& \textbf{Case 1} &&&&& \textbf{Case 2} &&\\\toprule
 method & type & \textbf{none} & $z^{(1)}$ & $z^{(2)}$ & $z^{(3)}$ & $z^{(4)},\dots, z^{(10)}$ & \textbf{none} & $z^{(1)}$ & $z^{(2)}$ & $z^{(3)}$ & $z^{(4)},\dots, z^{(10)}$ \\\midrule
 mobEmax & unrestr. & 0.92 & 0.01 & 0.01 & 0.01 & 0.06 & 0.00 & 0.92 & 0.02 & 0.93 & 0.11 \\ 
 mobEmax & restr. & 0.92 & 0.01 & 0.01 & 0.01 & 0.06 & 0.80 & 0.08 & 0.01 & 0.09 & 0.07 \\ 
 mobMeans & unrestr & 0.92 & 0.01 & 0.01 & 0.01 & 0.06 & 0.00 & 0.82 & 0.01 & 0.82 & 0.04 \\ 
 mobMeans & restr. & 0.92 & 0.01 & 0.01 & 0.01 & 0.05 & 0.92 & 0.02 & 0.01 & 0.02 & 0.05 \\  
 mobSpline & unrestr & 0.93 & 0.01 & 0.01 & 0.01 & 0.05 & 0.00 & 0.87 & 0.01 & 0.87 & 0.05 \\ 
 mobSpline & restr. & 0.93 & 0.01 & 0.01 & 0.01 & 0.05 & 0.89 & 0.03 & 0.01 & 0.03 & 0.05 \\\bottomrule
\end{tabular}

\begin{tabular}{rr|ccccc|ccccc}
 & &&& \textbf{Case 3} &&&&& \textbf{Case 4} &&\\\toprule
 method & type & none & $\boldsymbol{z^{(1)}}$ & $\boldsymbol{z^{(2)}}$ & $z^{(3)}$ & $z^{(4)},\dots, z^{(10)}$ & none & $\boldsymbol{z^{(1)}}$ & $\boldsymbol{z^{(2)}}$ & $z^{(3)}$ & $z^{(4)},\dots, z^{(10)}$ \\\midrule
 mobEmax & unrestr. & 0.00 & 0.89 & 0.93 & 0.04 & 0.18 & 0.61 & 0.19 & 0.20 & 0.02 & 0.10 \\ 
 mobEmax & restr. & 0.13 & 0.58 & 0.62 & 0.04 & 0.19 & 0.76 & 0.09 & 0.10 & 0.02 & 0.09 \\ 
 mobMeans & unrestr. & 0.00 & 0.79 & 0.79 & 0.01 & 0.05 & 0.72 & 0.12 & 0.13 & 0.01 & 0.05 \\ 
 mobMeans & restr. & 0.35 & 0.35 & 0.36 & 0.01 & 0.06 & 0.86 & 0.05 & 0.05 & 0.01 & 0.05 \\  
 mobSpline & unrestr. & 0.00 & 0.83 & 0.85 & 0.01 & 0.05 & 0.68 & 0.15 & 0.15 & 0.01 & 0.05 \\ 
 mobSpline & restr. & 0.27 & 0.42 & 0.44 & 0.01 & 0.08 & 0.82 & 0.06 & 0.06 & 0.01 & 0.06 \\\bottomrule
\end{tabular}

\begin{tabular}{rr|ccccc}
 & &&& \textbf{Case 5} &&\\\toprule
 method & type & none & $\boldsymbol{z^{(1)}}$ & $\boldsymbol{z^{(2)}}$ & $z^{(3)}$ & $z^{(4)},\dots, z^{(10)}$ \\\midrule
 mobEmax & unrestr. & 0.00 & 0.97 & 0.96 & 0.74 & 0.18  \\ 
 mobEmax & restr. & 0.02 & 0.92 & 0.77 & 0.07 & 0.22 \\ 
 mobMeans & unrestr. & 0.00 & 0.99 & 0.85 & 0.53 & 0.04 \\ 
 mobMeans & restr. & 0.33 & 0.53 & 0.28 & 0.01 & 0.06 \\  
 mobSpline & unrestr. & 0.00 & 0.99 & 0.89 & 0.60 & 0.04 \\ 
 mobSpline & restr. & 0.23 & 0.65 & 0.36 & 0.01 & 0.08 \\\bottomrule
\end{tabular}
 \label{tab:id1}
\end{table}

In case 1 and 2 there are no covariates interacting with the
treatment.  For case 1, where there are no covariate effects whatsoever, all
models do not partition the data in more than 90$\%$ of cases 
and thus control the type I error at the nominal level.  In case
2 restricting the partitioning to the treatment effect parameters
makes a big difference. If the partitioning is restricted in this way
there is often no partitioning of the data, but still more often than
in case 1. When there is no restriction, the correct covariates $z^{(1)}$
and $z^{(3)}$ are found with high frequency.

Cases 3-5 include covariate interactions with the treatment.  In cases
3 and 5 the method picks up the predictive effects of $z^{(1)}$ and $z^{(2)}$
well.  If partitioning is not restricted then the purely prognostic
covariate $z^{(3)}$ is also correctly included.  In case 4 finding
predictive effects on $\theta_{2}$ seems to be more challenging. None
of the methods reliably partition over the predictive covariates $z^{(1)}$
and $z^{(2)}$. These covariates do appear in more trees than $z^{(3)}$, which
is just noise in this case, but more than 50$\%$ of the time a global
model is fitted.

When comparing the different models, mobEmax seems to show the best
performance overall, mobSpline and mobMeans seem to be more greatly
affected by restricting partitioning.  They both partition less often.

\FloatBarrier

\paragraph{Estimating quantities of interest}
\label{sec:results-est}
In dose-finding trials two important quantities of interest are the
treatment effect at specific dose levels, as well as the minimum
effective dose (MED), which is the smallest dose required to achieve a
relevant treatment effect over placebo. Being able to estimate both of
these precisely is therefore of great importance for a method used in
these settings. In the situation we consider for this article this is
especially challenging, since the patient populations can be
heterogeneous with regards to both of these quantities.

To evaluate the treatment effect estimation of our methods we estimate
the treatment effect individually for each patient at a sequence of dose
levels from 1 to 100 using increments of 1. We average the squared error
of these 100 estimates for each patient and then average over all
patients to obtain the MSE of the individual treatment effect estimates.
We compare these estimates as obtained through \textit{mob} with the estimates
from a global $E_{max}$ model fit for all patients. We consider $\sigma = 0.12$ 
and $\sigma = 0.24$ as standard deviations,
where the latter setting simulates a scenario, where any effects are 
small in relation to the error variance and therefore harder to detect.

\begin{figure}[h!]
\centering
\textbf{(i)}

\includegraphics[width=0.7\textwidth]{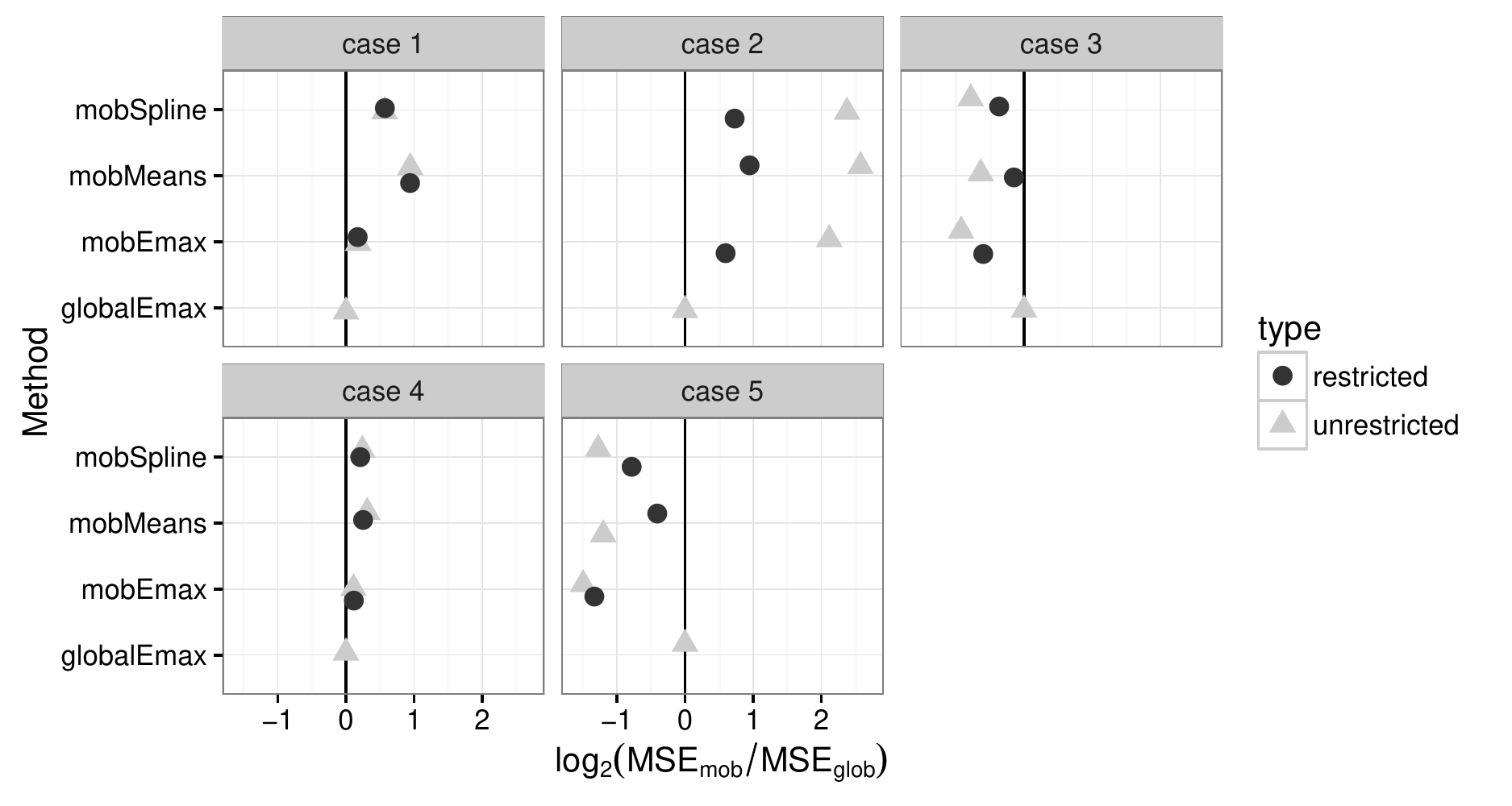}

\textbf{(ii)}

\includegraphics[width=0.7\textwidth]{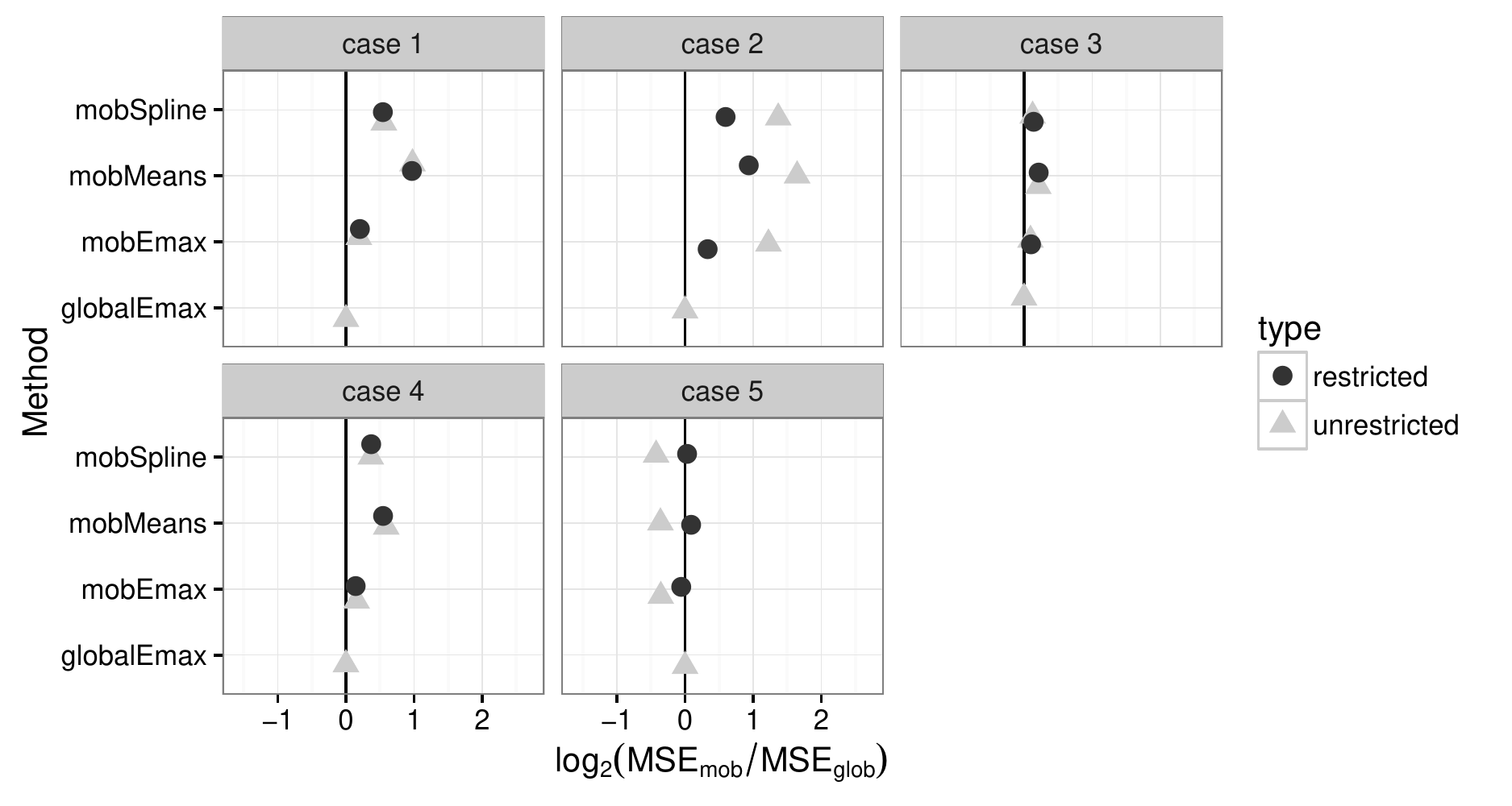}
\caption{Reduction of MSE for individual treatment effect estimation of the \textit{mob} methods compared to a global $E_{max}$ model
as $log_2(\frac{MSE_{mob}}{MSE_{globalEmax}})$ over 5000 simulations with
residual standard deviation of $\sigma= 0.12$ (i) and $\sigma= 0.24$ (ii).}
\label{fig:trt-est}
\end{figure}

Figure \ref{fig:trt-est}
shows the MSE of \textit{mob} with different dose-response models relative to the
global $E_{max}$ model.
In cases, where covariates do not affect the treatment effect (cases 1
and 2) \textit{mob} methods increase the MSE of treatment effect
estimates, since the global $E_{max}$ model is the better suited model
in these situations. When using \textit{mob} in combination with the
$E_{max}$ model, the increase is very small in case 1, since
\textit{mob} rarely partitions the data (see Section
\ref{sec:results-id}). For case 2 the effect of restricting splitting
to treatment effect parameters becomes visible.  Since covariate
effects here are only prognostic, restricting the parameters over
which to split, reduces the MSE. If splitting happens unrestricted,
MSEs are more than quadrupled in some situations, compared to the
global $E_{max}$ model.

In Cases 3 and 5 the advantages of partitioning the data become
clearly visible. Here using \textit{mob} methods leads to a clear reduction of
the MSEs for $\sigma = 0.12$, which are in some situations more than halved. 
For the larger $\sigma$ reduction can only be seen in Case 5, while in
Case 3 the MSEs are similar to those under the global model. 
Restriction of splitting on parameters relevant for the treatment effect weaken
the observed improvements somewhat. 

Case 4 seems to be a special situation, where there is no
big difference between the \textit{mob} methods and the global model. As also shown in
the results in Section \ref{sec:results-id}, covariate effects on the
$\theta_{2}$ parameter, which lead to steeper or shallower curves
seem to be much harder to identify.

Comparing the different \textit{mob} dose-response models to each
other, the correct $E_{max}$ model performs the best, having the
smallest MSEs of all \textit{mob} methods in most scenarios, as
expected as this is the true model. The remaining models, however,
show similar performance with slightly worse MSE in cases 3 and 5,
with the mobSpline typically being better than mobMeans.

\begin{figure}[h!]
\centering
\textbf{(i)}

\includegraphics[width=0.7\textwidth]{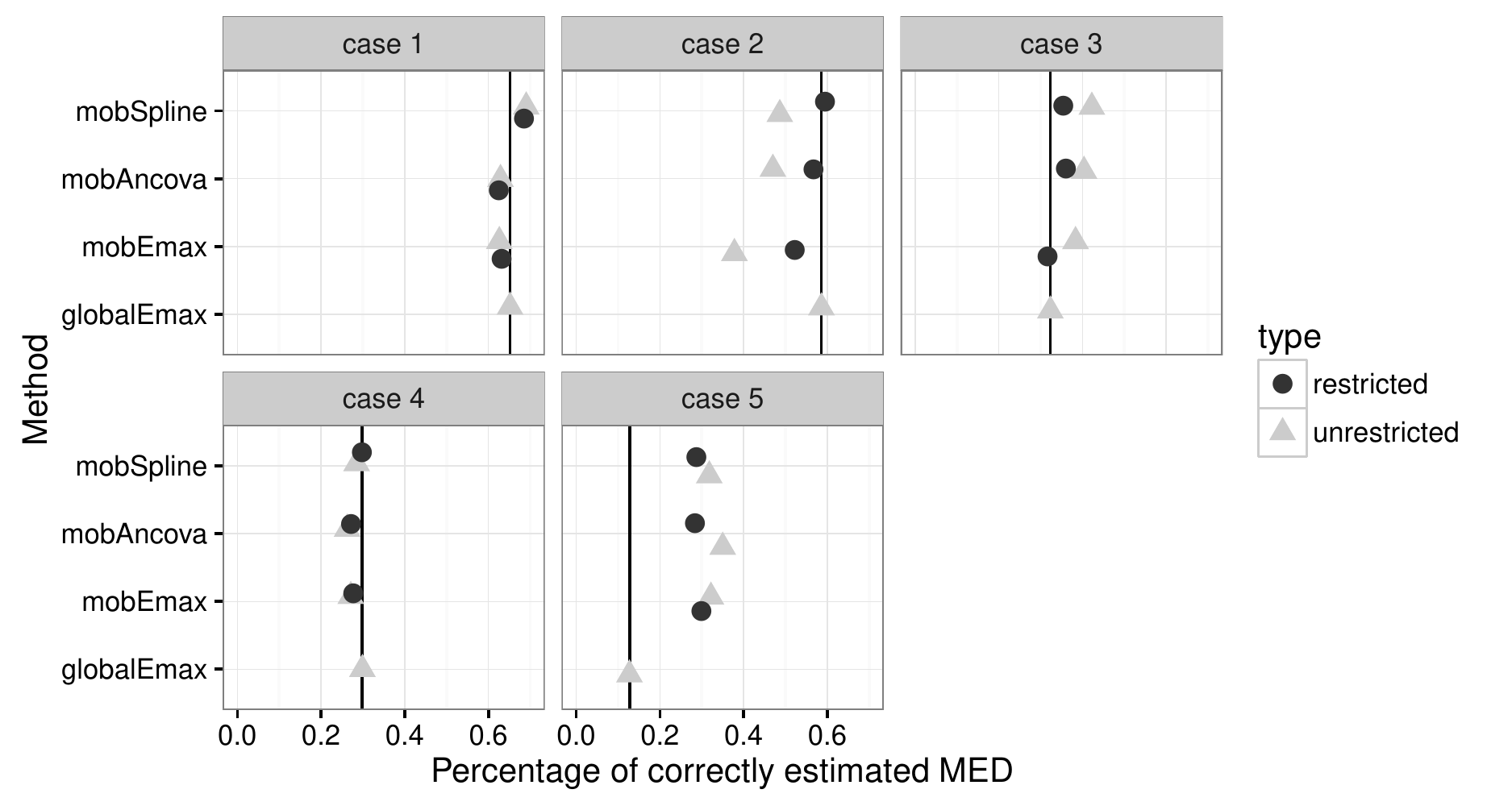}

\textbf{(ii)}

\includegraphics[width=0.7\textwidth]{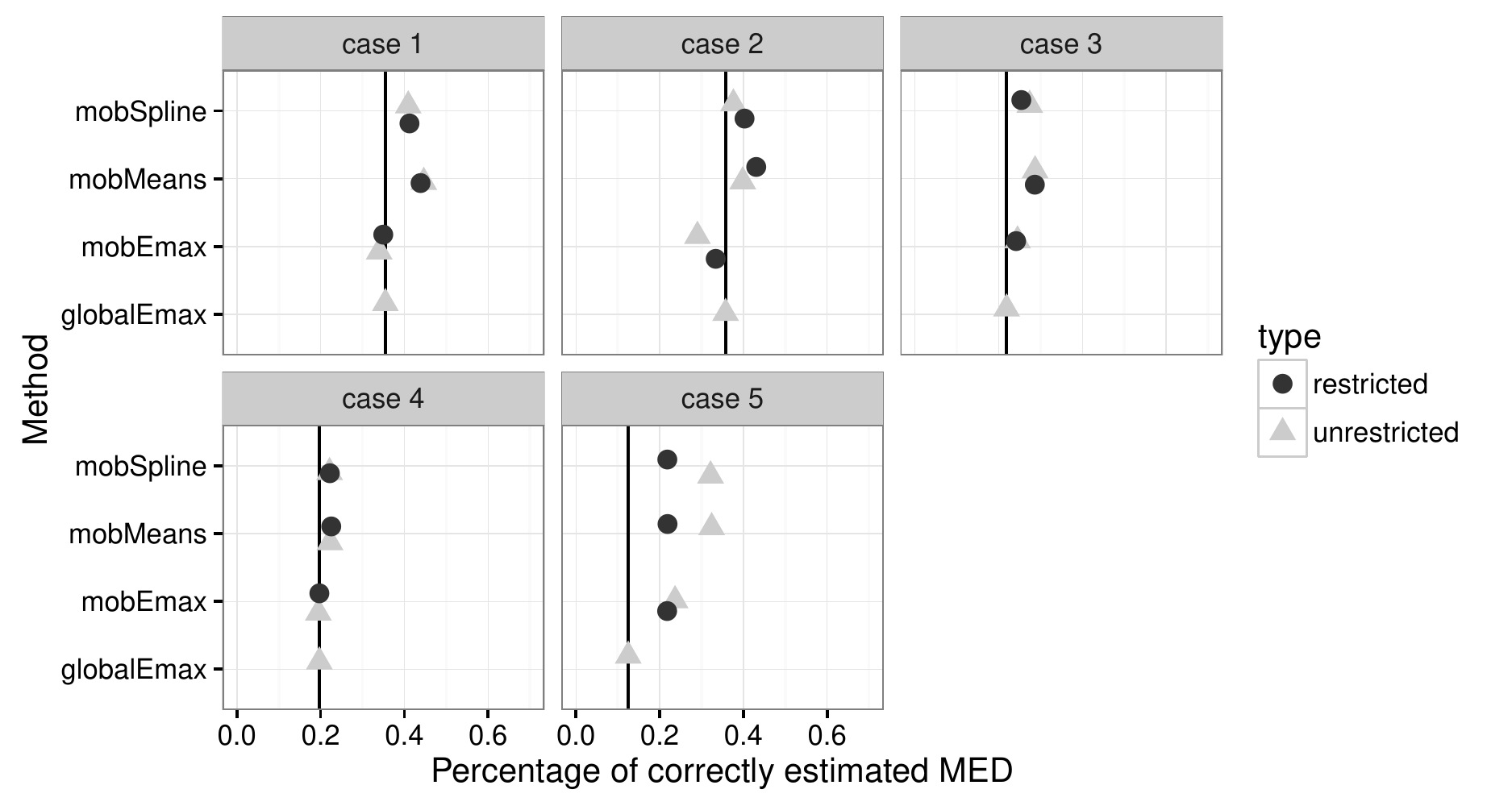}
\caption{Percentage of correctly estimated individual MED over 5000 simulations
for \textit{mob} methods and globalEmax model with
residual standard deviation of $\sigma= 0.12$ (i) and $\sigma= 0.24$ (ii).}
\label{fig:med-est}
\end{figure}

For the MED, we consider, that it is in general more important to
estimate MED accurately in areas of the dose-response curve, where the
curve is very steep. In these areas small deviations of the dose can
lead to large changes in response. We therefore assume that there is
an interval of doses around the true MED, which are all acceptable
doses to pick and the width of this interval depends on the steepness
of the curve. For our simulations we assume, that the minimum relevant
effect size is 0.1, but consider all doses $d$ for which
$\Delta(d) \in [0.08, 0.12]$ as acceptable choices for the MED. For
each simulated trial our metric is 1, if the estimated MED is in this
dose range and 0 otherwise. We then average this binary metric to
receive a percentage of correctly estimated MEDs over all simulated
trials. The results are displayed in Figure \ref{fig:med-est}.

Results for the MED are similar to that obtained for estimating the
individualized treatment effects (see Figure \ref{fig:med-est}).  MED
estimation is improved in cases 3 and 5, when using \textit{mob}.  For
a global model in case 5 only about 10$\%$ of estimated MEDs are in
the correct range, which shows the need for partitioning. Using
\textit{mob} around 30$\%$ of MED estimations are correct.  In the
remaining cases \textit{mob} performs usually slightly worse than the
global model. Case 2 again shows the effect of restricting to
parameters relevant for the treatment effect. In case 2 and for
$\sigma = 0.12$ unrestricted splitting has around 10 -15$\%$ fewer
correctly estimated MEDs compared to restricted splitting.  In general
the results are qualitatively similar for both considered error
variances, improvements in MED estimation can also be seen for the
high variance.

When comparing dose-response models, the $E_{max}$ model seems to have
less of an advantage for estimating MED compared to treatment effect
estimation.

\section{Subgroup analysis for the example trial}
\label{sec:ex-analysis}

We now return to the dose-finding trial described in Section
\ref{sec:motivating-example}.  Using the methods discussed above we
can perform an exploratory subgroup analysis.  We fit an $E_{max}$
model as described in Table \ref{tab:mod} and use the \textit{mob}
algorithm to search for subgroups.  We use control parameters of
\textit{alpha = 0.1, minsize = 20 and maxdepth = 4} and restrict the
partitioning to $\boldsymbol{\theta}$ to only identify subgroups,
which differ with regards to the treatment effects. The code to
perform the analysis can be found in the appendix.

The algorithm does partition the data and finds two subgroups of size
72 and 198 based on the binary variable $z^{(7)}$ with an unadjusted
$p$-value of 0.006 for the parameter instability test.  Table
\ref{tab:par_ex} shows the parameters in the global model and the
parameters of the $E_{max}$ model in the two subgroups as found by
\textit{mob}.  While the placebo effect stays roughly the same for all
3 models there are differences in $\theta_1$ and $\theta_2$, between
the two subgroups. For patients with $z^{(7)} = 1$ both $\theta_1$ and
$\theta_2$ are estimated to be much lower than for the remaining
patients.

\begin{figure}[h!]
\centering
\includegraphics[width=0.8\textwidth]{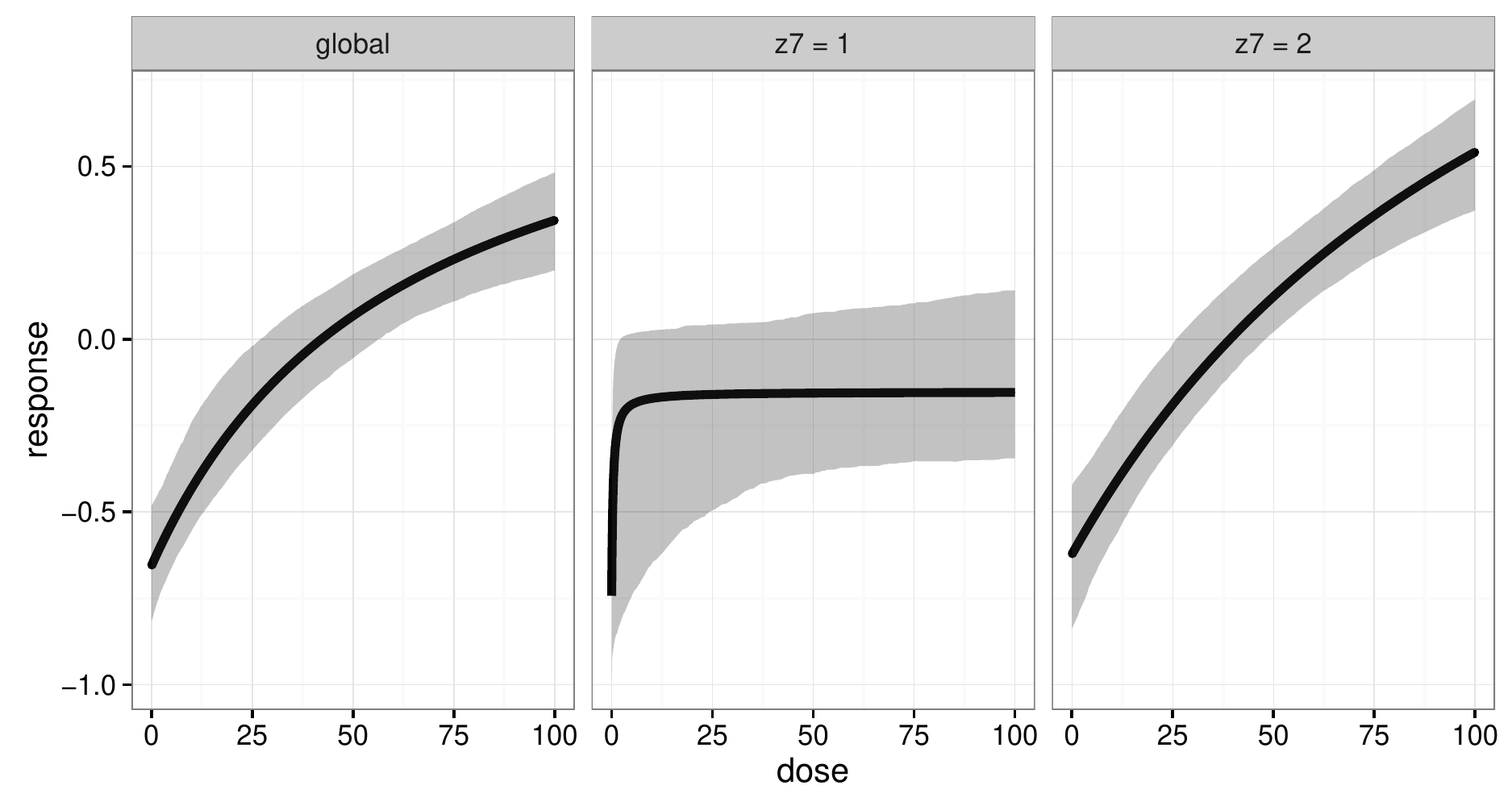}
\caption{Dose-response curves with 90$\%$ confidence bands 
for the global model and models in the
two subgroups, $z_7 = 1$ and $z_7 = 2$.}
\label{fig:mobcurves_ex}
\end{figure}

Figure \ref{fig:mobcurves_ex} shows the resulting dose-response curves
in the dose range. For the patients with $z^{(7)} = 1$ a plateau 
is reached very quickly and roughly half of the treatment effect of the other
group is achieved at the maximum dose level 100.
The curve of the patients with $z^{(7)} = 2$ is very similar to the
curve from the global model with slightly higher treatment effect
for higher dose levels.

To assess the robustness of these findings we repeated the analysis
with the other models shown in Table \ref{tab:mod}. Using these models
with the same settings, we did not detect a significant parameter
instability and thus the global model was fitted. Nevertheless, for
all models the split with the lowest $p$-value was the split over
covariate $z^{(7)}$ with unadjusted $p$-values 0.026 for mobMeans and
0.025 for mobSpline models. Therefore different models seem to be in
unison about which covariate causes most parameter instability. In
this context one should also consider the findings discussed in
Section \ref{sec:results-id}, where $E_{max}$ models detected existing
subgroups more often, while controlling the type I error as well as
other models. Of course, in these simulations the true underlying
model was also $E_{max}$. Nevertheless it could be the case, that
models other than the $E_{max}$ simply lack the power to detect this
covariate effect.

There seems to be some evidence that suggests an effect of the
covariate $z^{(7)}$ on the dose-response shape. The strong dose-response
relationship observed, when fitting a global model seems to be driven
by a subgroup of consisting of $\approx 73$\% of the patients. The
remaining patients on the other hand show a reduced treatment
effect.

\begin{table}[ht]
  \caption{Parameters of the $E_{max}$ model, fit globally
and in the subgroups as found by \textit{mob}.}
  \centering
   \begin{tabular}{ccccc}\toprule
   Group & size & $\theta_0$ & $\theta_1$ & $\theta_2$\\\midrule
    global & 270 & -0.656 & 1.619 & 61.895\\
    $z^{(7)}$ = 1 & 72 & -0.743 & 0.591 & 0.336 \\
    $z^{(7)}$ = 2 & 198 & -0.623 & 2.649 & 127.540\\\bottomrule
 \end{tabular}

 \label{tab:par_ex}
\end{table}

\newpage 

\section{Conclusions and discussion}
\label{sec:discussionconclusion}

In this article we discussed a strategy for subgroup identification in
dose-finding trials based on model-based recursive partitioning. 

The results depicted in Section~\ref{sec:results-fit} show that
\textit{mob} can be seen as an adaptive method, that implicitly tries
to balance between the bias introduced through omitting potential
covariate effects and the additional variance introduced by fitting
models to a partitioned data-set. This is also visible in Figures
\ref{fig:trt-est} and \ref{fig:med-est}. When using the correct
$E_{max}$ models (and when restricting to treatment effect parameters)
using \textit{mob} seems to have few disadvantages in regards to
estimation. When large effects of covariates are present using
\textit{mob} often leads to improvements in estimation of the
quantities of interest. In scenarios, where there are no covariate
effects or only prognostic ones, estimation is only slightly worse
than with the global model (at least when we restrict the splitting to
treatment effect parameters), since \textit{mob} will often simply fit
the global model in these situations (see Table \ref{tab:id1}).

From the simulation results in Section \ref{sec:simulation-study} it
becomes clear that for sample sizes, treatment effect sizes and error
variability commonly observed in clinical dose-finding trials,
identification of differential treatment effects is challenging. 
Due to the bias-variance trade-off, benefits of partitioning the data are
quickly diminishing as the error variance becomes larger, 
as one can see in Figures \ref{fig:ll-emax} and \ref{fig:trt-est}(ii). 
From this perspective the type I error control
is an important aspect of the method.  Our simulation results depicted
in Table \ref{tab:id1} show that type I error is indeed controlled in
scenarios without covariate effects. Thus, at the very least, the
chance of a false positive subgroup finding, which is one concern for
these types of analyses, is reduced.
 
The main challenge, when applying subgroup analyses in the context of
dose-response trials are that many models used in this context are
non-linear.  Apart from the non-linear $E_{max}$ model, we therefore
also considered models, which are linear in the parameters to compare
their performances. Even though we only considered data generated from
an $E_{max}$ model, the spline model showed generally good
performance, which was in most aspects only slightly worse than the
$E_{max}$ model. It could therefore be seen as an alternative to
non-linear models, as the linearity in the parameters might be
beneficial in some situations. Further research might be required to
assess the method's properties, when data is generated from other
models.

One advantage of the algorithm is that it allows to restrict the
partitioning over covariate effects to a specific set of
parameters. In the context of subgroup analyses this can be used to
distinguish between prognostic and predictive effects of covariates,
where the latter are usually of greater interest. The results in Table
\ref{tab:id1} show that these restrictions can effectively be used to
reduce splits over prognostic covariates. It is visible, that this
also reduces the detection rate of predictive effects, which was also
suggested in \cite{seib:zeil:hoth:2016}. This setting could therefore
be seen as a tuning parameter, similar to the other possible options
of the algorithm like the significance level, the minimum node size
and the maximum tree depth.

In this paper only normally distributed outcomes were considered for
the simulations and trial example, but an extension of this method to
other outcome types is possible. These extensions as well as the
methods discussed in this paper can be implemented using the R
package \textit{partykit}.

\section*{Acknowledgements}
The authors would like to thank Torsten Hothorn for helpful comments and discussions.
\small
 This work was supported by funding from the European Union's Horizon 2020 research and innovation programme under the Marie Sklodowska-Curie grant agreement No 633567 and by the Swiss State Secretariat for Education, Research and Innovation (SERI) under contract number 999754557 . The opinions expressed and arguments employed herein do not necessarily reflect the official views of the Swiss Government.\\

\begin{minipage}{.5\textwidth}
\includegraphics[scale = 0.03]{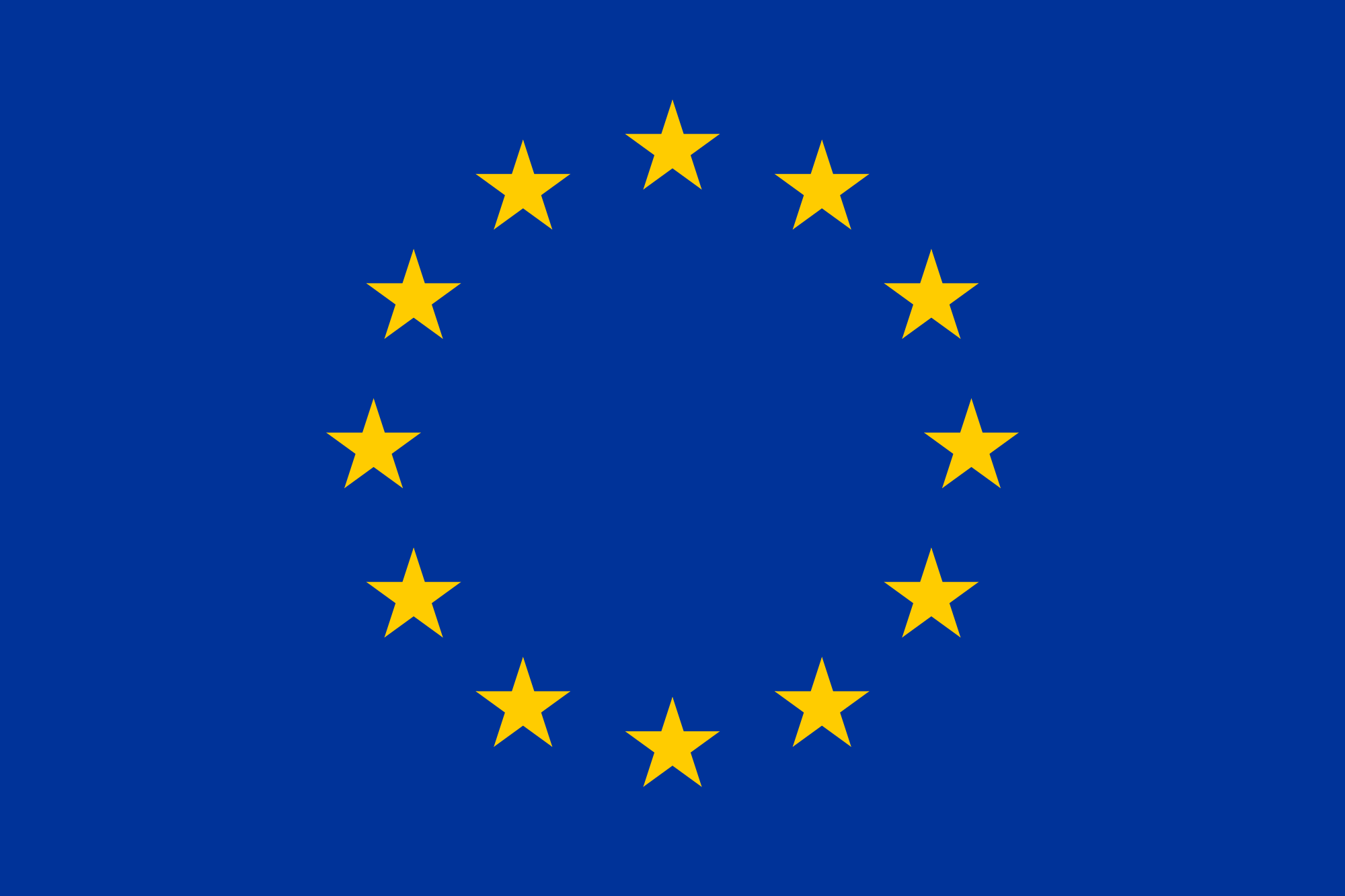}
\end{minipage}%
\begin{minipage}{.5\textwidth}
\includegraphics[scale = 1]{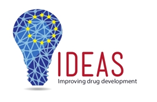}
\end{minipage}

\begin{appendices}
\section*{Appendix}
\subsection*{R code for trial example}
\scriptsize
\begin{lstlisting}
library(DoseFinding)
library(partykit)

# emax fitting function
emaxMob <- function(y, x, start = NULL, 
                          weights = NULL, 
                          offset = NULL,...,
                          estfun = FALSE, 
                          object = FALSE){
  model <- fitMod(resp = y, dose = x, model = "emax", 
  				start = start, bnds = defBnds(max(x))$emax, ...)
  sigma <- sqrt(model$RSS / model$df)
  coefficients <- coef(model)
  rss <- model$RSS
  if(estfun == 1){
    n.coef <- length(coef(model))
    param <- list(dose = x)
    param[2 : n.coef] <- as.list(coef(model)[-1])
    names(param)[2 : n.coef] <- names(coef(model))[-1]
    grad <- do.call(emaxGrad, param)
    fitted.y <- predict(model, predType= "full-model")
    res <- y - fitted.y
    scores <- -(res * grad)/sigma^2
  }
  else{
    scores <- NULL
  }
  if(object == 0)
    model <- NULL
  return(list(coefficients = coefficients, objfun = rss, 
              estfun = scores, object = model))
}
  
              
# mob analysis
partvar.names  <- paste(colnames(dr.dat)[1:10], collapse = "+")
form <- as.formula(paste("resp~0+dose|", partvar.names))
parm <- 2:3
alpha <- 0.1
minsize <- 20
tree <- mob(form,
    data=dr.dat,
    fit = emaxMob, 
    control = mob_control(bonferroni = TRUE, 
                          alpha = alpha, 
                          minsize = minsize,
                          parm = parm,
                          maxdepth = 4))

\end{lstlisting}

\end{appendices}

\bibliographystyle{wileyj}

\bibliography{bibl}
\end{document}